\documentclass[11pt]{article}
\usepackage[
  a4paper,
  top=26mm,
  bottom=26mm,
  left=26mm,
  right=26mm
]{geometry}

\usepackage{authblk}
\usepackage[numbers,sort&compress]{natbib}

\usepackage{xcolor}%
\usepackage{multirow}%
\usepackage{amsmath,amssymb,amsfonts}%
\usepackage{amsthm}%
\usepackage{mathabx}
\usepackage{diagbox}
\usepackage{dirtytalk}
\usepackage{enumitem}
\usepackage{hyperref}
\usepackage{cleveref} 
\usepackage{longtable}
\usepackage{mathtools}
\usepackage{tabularx}
\usepackage{thmtools}
\usepackage{placeins}
\usepackage{chngcntr}

\usepackage{hyperref}
\usepackage{url}
\hypersetup{
    colorlinks=true,
    urlcolor=blue,
    citecolor=blue,
    linkcolor=red
}

\declaretheoremstyle[
    headfont=\bfseries, 
    notefont=\bfseries, 
    notebraces={(}{)}, 
    bodyfont=\normalfont, 
]{boldstyle}

\declaretheorem[name=Theorem,style=boldstyle,numberwithin=subsection]{theorem}
\declaretheorem[name=Definition,style=boldstyle,numberlike=theorem]{definition}
\declaretheorem[name=Proposition,style=boldstyle,numberlike=theorem]{proposition}
\declaretheorem[name=Lemma,style=boldstyle,numberlike=theorem]{lemma}

\declaretheorem[name=Example,style=boldstyle,numberlike=theorem]{example}
\declaretheorem[name=Note,style=boldstyle,numberlike=theorem]{note}
\declaretheorem[name=Formulation,style=boldstyle,numbered=yes]{formulation}

\begin{document}

\newcommand{\indiff}{e}

\newcommand{\N}{\mathbb{N}}
\newcommand{\2}{\{0,1\}}
\newcommand{\3}{\{0,\indiff,1\}}

\newcommand{\A}{\mathcal{A}}
\newcommand{\B}{\mathcal{B}}

\newcommand{\Pref}{\textbf{Pref}(\A)}
\newcommand{\PrefVar}[1]{\textbf{Pref}(#1)}
\newcommand{\Pair}{\textbf{Pair}(N)}
\newcommand{\Prof}{\revision{\textbf{Weak}(\A)^N}}
\newcommand{\ProfVar}[1]{\textbf{Weak}(#1)^N}

\newcommand{\PrefStrict}{\revision{\textbf{StrictPref}(\A)}}
\newcommand{\PairStrict}{\revision{\textbf{StrictPair}(N)}}
\newcommand{\ProfileStrict}{\revision{\textbf{Strict}(\A)^N}}

\newcommand{\AltsSet}{\revision{\{a_1, a_2, a_3\}}}
\newcommand{\AltsSetGeneral}{\{a_1, a_2, a_3, \dots, a_A\}}
\newcommand{\NPairs}[1]{\binom{|#1|}{2}}
\newcommand{\Idx}{\A^{(2)}}

\newcommand{\VotesOne}[1]{\textbf{Votes}_1(#1)}
\newcommand{\AggregatesOne}[1]{\textbf{Aggregates}_1(#1)}

\newcommand{\Welfare}{w}
\newcommand{\WelfareSig}{\Welfare: \Prof \rightarrow \Pref}

\newcommand{\term}[1]{\textit{\textbf{#1}}}


\newcommand{\parsplit}{\par\vspace{6pt}}

\newenvironment{oneshot}[1]{\@begintheorem{#1}{\unskip}}{\@endtheorem}

\newcommand{\revision}[1]{#1}

\title{Arrow's Impossibility Theorem as a Generalisation of Condorcet's Paradox}

\renewcommand\Affilfont{\small}

\author[1,2]{Ori Livson \thanks{Corresponding author: ori.livson@sydney.edu.au}}
\author[1,2]{Mikhail Prokopenko}

\affil[1]{The Centre for Complex Systems,  University of Sydney, NSW 2006, Australia}
\affil[2]{School of Computer Science, Faculty of Engineering, University of Sydney, NSW 2006, Australia}

\date{}

\maketitle

\vspace{-18pt}

\begin{abstract}
    Arrow’s Impossibility Theorem is a seminal result of Social Choice Theory that demonstrates the impossibility of ranked-choice decision-making processes to jointly satisfy a number of intuitive and seemingly desirable constraints, \revision{including that aggregated preferences are transitive}. The theorem is often described as a generalisation of Condorcet's Paradox, wherein pairwise majority voting may fail to jointly satisfy the same constraints due to the occurrence of elections that result in \revision{intransitive} preference cycles. In the strict preference case, i.e., where indifference between alternatives is not allowed, D’Antoni’s \revision{formulation of Arrow's Impossibility Theorem makes this relationship explicit by constructing aggregations with preference cycles given Arrow's constraints other than transitivity}~\cite{dantoni}. In this paper, we generalise D'Antoni's methodology to apply to Arrow's Impossibility Theorem in full (i.e., accounting for weak preferences). \revision{We additionally leverage this generalisation to strengthen Arrow's Impossibility Theorem by establishing that the joint satisfaction of its constraints other than transitivity yields an aggregation consisting of an intransitive preference cycle that spans all available alternatives. Notably, a preference cycle that spans all alternatives precludes any global ranking information.}
\end{abstract}

\section{Introduction}

Arrow’s Impossibility Theorem is a seminal result of Social Choice Theory that demonstrates the impossibility of ranked-choice decision-making processes (i.e., social welfare functions) to jointly satisfy a number of intuitive and seemingly desirable constraints, i.e., Transitivity of Preferences, Unrestricted Domain, Unanimity, Independence of Irrelevant Alternatives (IIA) and Non-Dictatorship.

\parsplit

Intuition for Arrow's Impossibility Theorem originates in Condorcet's discovery that certain voting systems with 3 or more alternatives cannot guarantee \textit{majority rule}, i.e., the requirement that aggregate (i.e., winning) preferences are always shared by the majority of voters. Chief among examples of this is \textit{Condorcet's Paradox} on pairwise majority voting. In pairwise majority voting, certain elections fail to decide a winner due to the occurrence of a preference cycle, \revision{by which we mean} election outcomes where for alternatives $X$, $Y$ and $Z$: $X$ is preferred to $Y$, $Y$ to $Z$, and $Z$ to $X$\revision{, with at least one of those preferences being strict}.

\parsplit

Because pairwise majority voting satisfies all the constraints of Arrow’s Impossibility Theorem other than \revision{Transitivity of Preferences}, Arrow's Impossibility Theorem is often described as a generalisation of Condorcet's finding~\cite{sep-arrows-theorem}. \revision{This connection becomes particularly clear when Arrow's Impossibility Theorem is formulated as follows:}

\begin{formulation}\label{formulation}
    \revision{If a social welfare function satisfies Unrestricted Domain, Unanimity, Independence of Irrelevant Alternatives (IIA) and Non-Dictatorship there exists a profile of individual preferences whose aggregate is intransitive.}
\end{formulation}

\parsplit

\revision{However, direct proofs of Formulation 1 are lacking; the theorem is more typically proven by establishing that only dictatorial social welfare functions satisfy the remaining constraints.} \revision{Although recently, D'Antoni directly proved \revision{Formulation}~\ref{formulation} for the case where individual preferences are strict~\cite{dantoni}.} In this case, D'Antoni shows that an inherent problem in Condorcet's Paradox generalises to all social welfare functions satisfying \revision{Unrestricted Domain}, Unanimity, IIA and Non-Dictatorship.\footnote{We also refer the reader to~\cite{mossel,mossel-2}, which also proves Formulation 1 in the strict case using a related argument.}

\parsplit

The primary objective of our paper is to generalise D'Antoni's methodology to \revision{directly} prove \revision{Formulation}~\ref{formulation} fully --- i.e., \revision{allowing for indifference between alternatives (Theorem~\ref{theorem:arrow-full})}. The key step involves generalising D'Antoni's use of binary data (e.g., binary-valued tuples and matrices) for representing strict preferences, to ternary data for representing weak preferences.

\parsplit

\revision{Additionally, we are able to leverage this generalisation to prove a stronger result that guarantees the existence of profiles whose aggregate contains a preference cycle that spans all alternatives, rather than three in particular (Theorem~\ref{theorem:arrow-total-cycle}). This is significant because a cycle on all alternatives precludes any global ranking information (i.e., any notion of a first-ranked candidate, second-ranked candidate, and so on). By contrast, a preference cycle involving only three alternatives does not preclude the existence of another alternative that is preferred to all others.}

\section{Background}\label{section:background}

\subsection{Arrow's Impossibility Theorem}\label{subsection:background-arrow}

Social Choice Theory studies methods of aggregating individual inputs (e.g., votes, judgements, utility, etc.) into group outputs (e.g., election outcomes, sentencings, policies)~\cite{sep-social-choice}. Arrow's Impossibility Theorem concerns the aggregation of \term{weak orders}, i.e., transitive and complete relations. An example of \revision{a weak order} is a preferential voting ballot, wherein an individual vote is a ranking of alternatives from most to least preferred. In a weak order, tied rankings (i.e., \textit{indifference}) between alternatives \revision{are} permitted. We use the term \term{strict order} to refer to a weak order without indifference.

\parsplit

Following D'Antoni~\cite{dantoni}, a weak order on a fixed set of alternatives $\A$ can be represented by \revision{a binary relation $\preceq$ on $\A$ with strict part $\prec$ and symmetric part $\sim$, described in terms of preferences} as follows.
\begin{itemize}
    \item $a \preceq b$ for \revision{$b$ being \textit{weakly preferred} to $a$.}
    \item $a \prec b$ for \revision{$b$ being \textit{strictly preferred} to $a$  (i.e., $a \preceq b$ but $b \nprec a$).}
    \item $a \sim b$ for \textit{indifference} between $a$ and $b$ (i.e., $a \preceq b$ and $b \preceq a$).
\end{itemize}

Then, the defining properties for a weak order on alternatives $\A$ are correspondingly:
\begin{itemize}
    \item \textbf{Transitivity:} $\forall\ a,b,c \in \A$: if $a \preceq b$ and $b \preceq c$ then $a \preceq c$.

    \item \textbf{Completeness:} $\forall\ a,b \in \A$: one of $a \prec b$, $b \prec a$ or $a \sim b$ hold\revision{s}.
\end{itemize}

Moreover, weak orders may be written as a chain of the symbols $\prec$, $\sim$ and $\preceq$. For example, If $\A = \{a,b,c\}$, $a \prec b \sim c$ denotes the weak order consisting of $a \prec b$, $b \sim c$ and $a \prec c$ (by transitivity).

\parsplit

Given a fixed number $N \in \N$ of individuals, a \term{profile} is an $N$-tuple of weak orders. An example of a profile is an election, i.e., a tuple containing a single ballot from each individual. Note, each individual has a fixed index in the tuple across profiles. \revision{Then, for the purposes of this paper,} a \term{social welfare function} is a function from a set of \textit{valid} profiles to a \revision{complete binary relation on the alternatives}. Invalid profiles are those that would otherwise fail to aggregate to a \revision{complete relation}.

\parsplit

\begin{definition}\label{definition:fairness-conditions}
    A social welfare function satisfies:
    \begin{itemize}
        \item \textbf{Unrestricted Domain}: If all profiles are valid, i.e., can be aggregated.

        \item \textbf{Unanimity}: If \revision{when} all individuals share a strict preference of $a$ over $b$ the aggregate does too.

        \item \textbf{Independence of Irrelevant Alternatives (IIA)}: The outcome of aggregation with respect to alternatives $a$ and $b$ only depends on the individual preferences with respect to $a$ and $b$.

        \item \textbf{Non-Dictatorship}: There is no individual that irrespective of the profile, their strict preferences are always present in the aggregate outcome.

        \item \revision{\textbf{Transitivity}: If aggregation always produces a transitive relation.}
    \end{itemize}
\end{definition}

\parsplit

\begin{theorem}[Arrow's Impossibility Theorem]\label{theorem:arrows-impossibility-theorem-orig}
    \revision{No social welfare function on at least 3 alternatives and at least 2 individuals can simultaneously satisfy Unrestricted Domain, Unanimity and IIA,  Non-Dictatorship and Transitivity.}
\end{theorem}

\noindent For examples of proofs of Arrow's Impossibility Theorem, see~\cite{geanakoplos,arrow-one-shot}.

\subsection{Condorcet's Paradox}\label{subsection:condorcet-paradox}

Condorcet's Paradox refers to phenomena where a voting system on 3 or more alternatives cannot guarantee that winners that are always preferred by a majority of voters. A canonical example of a Condorcet Paradox is the observation that for the profile specified by Table~\ref{table:condorcet-plain}, pairwise majority voting cannot decide a winner \revision{as it} aggregates to a preference cycle.

\begin{table}[h!]
    \centering
    \begin{tabularx}{0.95\textwidth}{|p{1.8in}|X|X|X|}
        \hline
        \diagbox{Ranking}{Individual} & \textbf{1} & \textbf{2} & \textbf{3} \\\hline
        \textbf{1}                    & a          & b          & c          \\\hline
        \textbf{2}                    & b          & c          & a          \\\hline
        \textbf{3}                    & c          & a          & b          \\\hline
    \end{tabularx}
    \caption{A Profile on 3 voters and 3 alternatives $\{a,b,c\}$ that under pairwise majority voting, produces a Condorcet Paradox.}
    \label{table:condorcet-plain}
\end{table}

Pairwise majority voting is defined by ranking alternatives $x \prec y$ if more voters strictly prefer $x$ to $y$ than $y$ to $x$, and $x \sim y$ if there is a tie. If we apply this rule to the profile defined by Table~\ref{table:condorcet-plain}, we find that the majority of individuals strictly prefer $a$ to $b$ (individuals 1 and 3) as well as $b$ to $c$ (individuals 1 and 2), and $c$ to $a$ (individuals 2 and 3). Thus, aggregation yields a preference cycle $a \prec b \prec c \prec a$, which is contradictory given the requirement that aggregated preferences are transitive.

\parsplit

It is a simple exercise to verify pairwise majority voting satisfies \revision{Unrestricted Domain}, Unanimity, IIA and Non-Dictatorship, but as we have seen \revision{not Transitivity}.

\subsection{D'Antoni's Approach}\label{subsection:dantoni-background}

D'Antoni \revision{directly} established that, for strict orders, all social welfare functions satisfying \revision{Unrestricted Domain}, Unanimity, IIA and Non-Dictatorship violate \revision{Transitivity} by necessarily producing a profile \revision{whose aggregate contains a preference cycle}~\cite{dantoni}. Their methodology begins with the definition of a class of objects that encompasses both strict orders and preference cycles.  For example, for the 3 alternative case, indexed arbitrarily, say, $a_1, a_2, a_3$: the objects are tuples $(t_{\revision{12}}, t_{\revision{23}}, t_{\revision{31}})$, where $t_{\revision{12}}, t_{\revision{23}}, t_{\revision{31}}$ range over $\{0, 1\}$. For a strict order $\prec$ on $\{a_1, a_2, a_3\}$:
\begin{align*}
    t_{\revision{12}} = 0 \iff a_1 \prec a_2 &  & t_{\revision{23}} = 0 \iff a_2 \prec a_3 &  & t_{\revision{31}} = 0 \iff a_3 \prec a_1 \\
    t_{\revision{12}} = 1 \iff a_2 \prec a_1 &  & t_{\revision{23}} = 1 \iff a_3 \prec a_2 &  & t_{\revision{31}} = 1 \iff a_1 \prec a_3
\end{align*}

There are $2^3 = 8$ possible binary 3-tuples on $\{a_1,a_2,a_3\}$. This includes all 6 possible strict orders on 3 alternatives (see Equation~\ref{equation:correspondence}), and the tuples $(0, 0, 0)$ and $(1, 1, 1)$, which correspond to the preference cycles $\revision{a_1} \prec a_2 \prec a_3 \prec a_1$ and $a_3 \prec a_2 \prec a_1 \prec a_3$.

\begin{equation}
    \begin{array}{lll}
        (0,0,1) & a_1 \prec a_2 \prec a_3 & \qquad\qquad\qquad (0,1,0) \quad a_3 \prec a_1 \prec a_2  \\
        (0,1,1) & a_1 \prec a_3 \prec a_2 & \qquad\qquad\qquad (1,0,0) \quad a_2 \prec a_3 \prec a_1  \\
        (1,0,1) & a_2 \prec a_1 \prec a_3 & \qquad\qquad\qquad (1,1,0) \quad a_3 \prec a_2 \prec a_1.
    \end{array}
    \label{equation:correspondence}
\end{equation}

\begin{table}[h!]
    \centering
    \begin{tabularx}{0.95\textwidth}{|p{1.8in}|X|X|X|}
        \hline
        \diagbox{Alternative}{Individual} & \textbf{1} & \textbf{2} & \textbf{3} \\\hline
        \textbf{$a_1$} (vs $a_2$)         & 0          & 1          & 0          \\\hline
        \textbf{$a_2$} (vs $a_3$)         & 0          & 0          & 1          \\\hline
        \textbf{$a_3$} (vs $a_1$)         & 1          & 0          & 0          \\\hline
    \end{tabularx}
    \caption{The Condorcet profile of Section~\ref{subsection:condorcet-paradox} Table~\ref{table:condorcet-plain}, written with alternatives $a_1,a_2,a_3$ in place of $a,b,c$ respectively\revision{.}}
    \label{table:dantoni-profile}
\end{table}

An $N$ individual profile on the 3 alternatives can then be represented by a binary-valued $3 \times N$ matrix (see Table~\ref{table:dantoni-profile} for an example). The rows of these matrices record each individual's preferences on a single pair of alternatives, and each column records an individual's preferences. D'Antoni proceeds to use binary valued data such as this to \revision{prove} Arrow's Impossibility Theorem for strict preferences.

\section{Framework}\label{section:framework}

\subsection{Weak Orders and Preference Cycles}

The key generalising step of this paper with respect to  D'Antoni's framework in~\cite{dantoni} is the use of ternary data to represent weak preferences, i.e., for alternatives $x$ and $y$ using 0 to represent $x \prec y$, 1 for $y \prec x$ and a third value $\indiff$ for indifference: $x \sim y$. \revision{Another generalisation that we make explicit is the definition of this representation for over 3 alternatives. While IIA implies that proving \revision{Formulation}~\ref{formulation} in the 3 alternative case yields Arrow's Impossibility Theorem for additional alternatives~\cite[Footnote 3]{dantoni}, this generalisation is needed for deriving additional results in Section~\ref{section:more}.}

\parsplit

\revision{To begin, recall that a complete binary relation $\preceq$ on $\A$ is a weak order when it is additionally transitive. Conversely, we define a preference cycle on a triple of alternatives $a,b,c \in \A$ as a chain of preferences $a \preceq b \preceq c \preceq a$ where at least one comparison is strict. In fact, a complete relation contains a preference cycle if and only if it is intransitive (see Proposition~\ref{proposition:complete-order-classification}). When a complete relation contains a preference cycle that spans all alternatives, we say that it \textit{is} a preference cycle. Otherwise, we only say that the relation contains a preference cycle.

    \parsplit

    Next, recall D'Antoni's representation of strict, complete relations on 3 alternatives as binary-valued tuples of length 3. To generalise this representation, i.e., to allow for additional alternatives and indifference, we use ternary-valued tuples of length given by the binomial coefficient $\binom{|\A|}{2}$. We begin with the simpler 3 alternative case of this representation, which generalises D'Antoni's 3 alternative definition in~\cite[Section 2]{dantoni}.
}

\parsplit

\begin{definition}[Preference Relations on 3 Alternatives]\label{definition:abstract-pref}
    \revision{Let $\A \coloneqq \AltsSet$ be a set of 3 alternatives} and $\{0, \indiff, 1\}$ a set of ternary values. A \term{preference relation} on $\A$ is a ternary valued \revision{tuple of length 3, i.e., a tuple $(t_{12}, t_{23}, t_{31})$} with each $t_{\revision{xy}} \in \3$.

    \parsplit

    Moreover, every preference relation \revision{$(t_{12}, t_{23}, t_{31})$, corresponds to the following complete relation $\preceq$ on $\A$, where for every $xy \in \{12, 23, 31\}$\textnormal{:}}
    \begin{equation*}
        t_{\revision{xy}} = \begin{cases}
            \indiff \iff a_\revision{x} \sim a_\revision{y} \\
            0 \iff a_\revision{x} \prec a_\revision{y}      \\
            1 \iff a_\revision{x} \succ a_\revision{y}
        \end{cases}
    \end{equation*}
\end{definition}

\parsplit

\begin{example}\label{example:order-sequences}
    For $ \A \coloneqq \{a_1, a_2, a_3\}$, the weak order $a_1 \sim a_2 \prec a_3$ corresponds to $(\indiff, 0, 1)$ and $a_3 \sim a_1 \prec a_2$ to $(0, 1, \indiff)$. The  preference cycle $a_1 \prec a_2 \prec a_3 \sim a_1$ can be written as $(0,0,\indiff)$.
\end{example}

\revision{In general, to represent a complete relation on $\A = \AltsSetGeneral$ by a ternary-valued tuple, we require that for each pair of alternatives $a,b \in \A$, exactly one of the tuple's values records whether $a \prec b$, $b \prec a$ or $a \sim b$. To do so, we introduce the following useful notation. Let $\Idx$ be the following set of strings:

    \begin{equation*}
        \Idx \coloneqq \{ij \mid i,j \in \N \text{ such that } 1 \leq i < j \leq A\}\ \setminus\ \{1A\} \cup \{A1\}
    \end{equation*}

    \parsplit

    For example, when $A = 3$ we have that $\Idx = \{12,23,31\}$ and when $A = 4$ we have that $\Idx = \{12,13,23,24,34,41\}$. Replacing $1A$ with $A1$ ensures that the following definition of preference cycles generalises Definition~\ref{definition:abstract-pref}, and will also be useful for determining when complete relations have preference cycles.

    \parsplit

    \begin{definition}[Preference Relations, General Case]\label{definition:abstract-pref-general}
        Let $\A \coloneqq \AltsSetGeneral$ be a set of 3 or more alternatives. A \term{preference relation} on $\A$ is a tuple $(t_{xy})_{xy\, \in \Idx}$ with each $t_{xy} \in \3$.

        \parsplit

        Moreover, every preference relation $(t_{xy})_{xy\, \in \Idx}$ corresponds to the complete relation $\preceq$ on $\A$, where for each $xy \in \Idx$:
        \begin{equation*}
            t_{xy} = \begin{cases}
                \indiff \iff a_x \sim a_y \\
                0 \iff a_x \prec a_y      \\
                1 \iff a_x \succ a_y
            \end{cases}
        \end{equation*}
        \indent We denote the set of all preference relations on alternatives $\A$ as $\Pref$.
    \end{definition}

    \noindent Note, a preference relation of all 0's corresponds to the preference cycle $a_1 \prec a_2 \prec a_3 \prec \dots \prec a_A \prec a_1$, and $a_1 \succ a_2 \succ a_3 \succ \dots \succ a_A \succ a_1$ given all 1's (see Proposition~\ref{proposition:spanning-cycles}).
}

\parsplit

We conclude this section by defining a negation operation on ternary data, and a condition which delineates when preference-relations correspond to weak orders vs \revision{complete relations with} preference cycles.

\parsplit

\parsplit

\begin{definition}\label{definition:neg}
    We define $\neg: \3 \rightarrow \3$ by the mappings $0 \mapsto 1$, $1 \mapsto 0$ and $\indiff \mapsto \indiff$. \revision{Then, for $(t_{xy})_{xy\, \in \Idx} \in \Pref$ and any $ab \in \Idx$, we write $t_{ba} \coloneqq \neg t_{ab}$.}
\end{definition}

\parsplit

\begin{restatable}{lemma}{test}\label{lemma:condorcet-condition}
    Let $\A \coloneqq \AltsSetGeneral$. The preference relation $t = \revision{(t_{xy})_{xy\, \in \Idx}}$ corresponds to a weak order if and only if \revision{$\forall xy, yz \in \Idx$}, either of the following hold\revision{s}. \revision{\begin{enumerate}
            \item $t_{xy} = t_{yz} = t_{zx} = \indiff$
            \item $\{0, 1\} \subseteq \{t_{xy}, t_{yz}, t_{zx}\}$
        \end{enumerate}}

    \noindent \revision{where by Definition~\ref{definition:neg}, when $zx \neq A1$, $t_{zx} \coloneqq \neg t_{xz}$. Furthermore, if neither 1 nor 2 is satisfied for some $xy, yz \in \Idx$}, $t$ corresponds to a \revision{complete relation with a} preference cycle.
\end{restatable}

\subsection{Profiles}

As in Section~\ref{subsection:dantoni-background}, a profile is a ternary-valued matrix, where the rows record each individual's preferences on a single pair of alternatives, and each column records a single individual's preferences.

\parsplit

\begin{definition}[Profiles and Pairwise Preferences]\label{definition:profiles}
    Let $\A \coloneqq \AltsSetGeneral$ be a set of alternatives, $N \geq 2$ be a number of individuals, and $m$ a $\3$-valued $\revision{\NPairs{\A}} \times N$ matrix. We define the following \textit{tuple-of-tuple} representations of $m$:
    \begin{itemize}
        \item A tuple of columns $(c_1, c_2, \dots, c_N)$, with each $c_i \in \Pref$ representing an individual's preferences. The matrix $m$ \revision{is} a \term{profile} if and only if every $c_i$ is a weak order. We denote the set of all profiles as $\Prof$.

              \parsplit

        \item A tuple of rows $\revision{(r_{xy})_{xy\, \in \Idx}}$, each representing the \term{pairwise preferences} across individuals, e.g., $\revision{r_{23}}$ contains each individual's preferences with respect to $a_2$ vs $a_3$. We denote the set of all possible rows by $\Pair$.
    \end{itemize}
\end{definition}

\noindent \revision{We then define functions $\neg: \Pair \rightarrow \Pair$ and $\neg: \Prof \rightarrow \Prof$ by applying $\neg: \3 \rightarrow \3$ coordinate-wise.}

\parsplit

\begin{example}\label{example:profile}
    A profile $m$ on $N = 4$ individuals and alternatives $\A = \{a_1, a_2, a_3\}$ is detailed by the following table.

    \begin{center}
        \begin{tabularx}{0.95\textwidth}{|p{1.8in}|X|X|X|X|}
            \hline
            \diagbox{Alternative\revision{s}}{Individual} & \textbf{1} & \textbf{2} & \textbf{3} & \textbf{4} \\\hline
            \textbf{$a_1$ vs $a_2$}                       & $\indiff$  & $\indiff$  & 0          & 0          \\\hline
            \textbf{$a_2$ vs $a_3$}                       & 0          & $\indiff$  & 1          & 0          \\\hline
            \textbf{$a_3$ vs $a_1$}                       & 1          & $\indiff$  & 1          & 1          \\\hline
        \end{tabularx}
    \end{center}

    The column form $(c_1, c_2, c_3, c_4)$ of $m$ comprises the following individual preference relations.
    \begin{center}
        {\normalsize
            \begin{tabular}{c c c}
                \textbf{Column}                 & \textbf{Individual Preference Relation} \\
                \hline
                $c_1=(\indiff,0,1)$             & $a_1 \sim a_2 \prec a_3$                \\
                $c_2=(\indiff,\indiff,\indiff)$ & $a_1 \sim a_2 \sim a_3$                 \\
                $c_3=(0,1,1)$                   & $a_1 \prec a_3 \prec a_2$               \\
                $c_4=(0,0,1)$                   & $a_1 \prec a_2 \prec a_3$               \\
            \end{tabular}}
    \end{center}

    Likewise, the row form $(r_{\revision{12}}, r_{\revision{23}}, r_{\revision{31}})$ of $m$, corresponds to the following pairwise preferences:

    \begin{center}
        {\normalsize
            \begin{tabular}{c c c c c}
                \textbf{Row}                                & \textbf{Individual 1} & \textbf{Individual 2} & \textbf{Individual 3} & \textbf{Individual 4} \\
                \hline
                $r_{\revision{12}} = (\indiff,\indiff,0,0)$ & $a_1 \sim a_2$        & $a_1 \sim a_2$        & $a_1 \prec a_2$       & $a_1 \prec a_2$       \\
                $r_{\revision{23}} = (0,\indiff,1,0)$       & $a_2 \prec a_3$       & $a_2 \sim a_3$        & $a_2 \succ a_3$       & $a_2 \prec a_3$       \\
                $r_{\revision{31}} = (1,\indiff,1,1)$       & $a_3 \succ a_1$       & $a_3 \sim a_1$        & $a_3 \succ a_1$       & $a_3 \succ a_1$       \\
            \end{tabular}}
    \end{center}
\end{example}

\parsplit

\begin{definition}[Strict Subsets]
    Given the sets $\Prof$, $\Pref$ and $\Pair$ of profiles, preference relations and pairwise preferences of alternatives $\A$ and $N \geq 2$ individuals, we write $\ProfileStrict$, $\PrefStrict$ and \revision{$\PairStrict$} for their strict (i.e., $\2$-valued) subsets, respectively.
\end{definition}

\revision{\begin{note}\label{note:profile-test}
        In the 3 alternative case, a matrix is a profile if and only if each of its columns is a weak order. By Lemma~\ref{lemma:condorcet-condition}, this occurs when each column either has a $0$ and a $1$ in it, or consists entirely of $\indiff$. For example, for $r \in \Pair$ and $r' \in \PairStrict$, the matrix in row form $(r, r', \neg r')$ must be a profile because each column has a $0$ or $1$ in row $r'$ and correspondingly a $1$ or $0$ in the same places in $\neg r'$.
    \end{note}
}

\subsection{Social Welfare Functions}

\parsplit

\begin{definition}\label{definition:generalised-social-welfare-function}
    A \term{social welfare function} on a set $\A = \AltsSetGeneral$ of 3 or more alternatives and $N \geq 2$ individuals is a function from a set of profiles to preference relations, i.e., a function \revision{$\Welfare: \mathcal{D} \rightarrow \Pref$ for $\mathcal{D} \subseteq \Prof$.}
\end{definition}

\noindent\revision{While $\Pref$ is defined with respect to an indexing (i.e., permutation) of the elements of $\A$, a social welfare function of the above kind always defines a unique mapping of $N$-tuples of weak orders to complete relations. Hence, the remaining definitions of this section are} generalisations of the constraints of Arrow's Impossibility Theorem, and are equivalent to D'Antoni's (see~\cite[Section 3]{dantoni}) on strict profiles.

\revision{\parsplit

    \begin{definition}\label{definition:quasi-unrestricted-domain}
        A social welfare function $\Welfare: \mathcal{D} \rightarrow \Pref$ satisfies:
        \begin{itemize}
            \item \term{Transitivity} when the \revision{image of $\Welfare$} only contains weak orders.

            \item \term{Unrestricted Domain} when $\mathcal{D} = \Prof$, i.e., aggregation at minimum produces a complete relation.
        \end{itemize}
    \end{definition}

    \begin{note}\label{note:equivalence}
        To simplify our remaining generalisations of Arrow's conditions (i.e., IIA, Unrestricted Domain and Non-Dictatorship), we will restrict our definition of IIA to social welfare functions that satisfy Unrestricted Domain, and then restrict our definition of Unanimity and Non-Dictatorship to those that satisfy IIA. This approach follows D'Antoni~\cite{dantoni}, and suffices for proving Arrow's Impossibility Theorem due to the logical equivalence of the following two statements.

        \begin{enumerate}
            \item No Social Welfare Function satisfies all of Unrestricted Domain, IIA, Unanimity, Non-Dictatorship and Transitivity, i.e., the standard form of Arrow's Impossibility Theorem.

            \item If a Social Welfare Function satisfies Unrestricted Domain, IIA, Unanimity and Non-Dictatorship then it violates Transitivity (Formulation 1 \textnormal{\&} Theorem~\ref{theorem:arrow-full}).
        \end{enumerate}

    \end{note}

}

\noindent Recall that IIA is the requirement that the aggregated preference of alternatives $X$ vs $Y$ depends only on the individual preferences regarding $X$ vs $Y$. Thus, because pairwise comparisons are given at the row-level, IIA is equivalent to the statement that a social welfare function can be decomposed into \textit{row-wise} functions.

\parsplit

\begin{definition}[IIA]\label{definition:iia}
    Let $\WelfareSig$ be a social welfare function \revision{that satisfies Unrestricted Domain.} The social welfare function $\Welfare$ satisfies \term{Independence of Irrelevant Alternatives (IIA)} when it can be expressed \revision{as a product of} functions $\revision{(s_{xy}: \Pair \rightarrow \3)_{xy\, \in \Idx}}$. Specifically, for every profile in row form \revision{$(r_{xy})_{xy\, \in \Idx}$} we have that:
    \begin{equation*}
        \revision{\Welfare((r_{xy})_{xy\, \in \Idx}) = (s_{xy}(r_{xy}))_{xy\, \in \Idx}}
    \end{equation*}
    \indent We call $\revision{(s_{xy})_{xy\, \in \Idx}}$ the \term{pairwise comparison functions} of $\Welfare$. \revision{Moreover, as in Definition~\ref{definition:neg}, for $xy \in \Idx$ and $r \in \Pair$, we denote $s_{yx}(r) \coloneqq \neg s_{xy}(\neg r)$.}
\end{definition}

\revision{

    \parsplit

    \begin{example}\label{exmaple:pairwise-swf}
        Pairwise majority voting corresponds to the following social welfare function $\WelfareSig$ that satisfies IIA. To begin, for $v \in \{0,1\}$ and $r = (u_1, u_2, \dots, u_N) \in \Pair$, we define $\textbf{Votes}_v(r) \coloneqq \{i \in \{1,2,\dots,N\} \mid u_i = v\}$. For instance, if $r$ records each individual's pairwise preferences on alternatives $a$ vs $b$, $\textbf{Votes}_0(r)$ records the subset of individuals who preference $a \prec b$ in $r$. Thus, $\Welfare$ can be defined as the product of the following pairwise comparison functions $\revision{(s_{xy}: \Pair \rightarrow \3)_{xy\, \in \Idx}}$, where for each $xy \in \Idx$ and $r \in \Pair$:
        \begin{equation}\label{equation:pcf}
            s_{xy}(r) = \begin{cases}
                0       & \text{ when } |\textbf{Votes}_0(r)| > |\textbf{Votes}_1(r)| \\
                1       & \text{ when } |\textbf{Votes}_0(r)| < |\textbf{Votes}_1(r)| \\
                \indiff & \text{ when } |\textbf{Votes}_0(r)| = |\textbf{Votes}_1(r)|
            \end{cases}
        \end{equation}
        \indent An example of Condorcet's Paradox then arises out of the aggregation of the profile $m \in \ProfVar{\{a_1,a_2,a_3\}}$ of Table~\ref{table:dantoni-profile} due to the following calculation:
        \begin{equation*}
            \Welfare(m) = (s_{12}(0,1,0),\ s_{23}(0,0,1),\ s_{31}(1,0,0)) = (0,0,0)
        \end{equation*}
        \indent Indeed, the tuple $(0,0,0) \in \Pref$ corresponds to a preference cycle on $a_1$, $a_2$ and $a_3$ by Lemma~\ref{lemma:condorcet-condition}.

        \parsplit

        \indent Note, that to define pairwise majority voting, the same pairwise comparison function was used for every pair of alternatives; this property is known as \say{neutrality}. An example of a non-neutral variant of $\Welfare$ involves replacing the last case of Equation~\ref{equation:pcf} with a tie-breaker that depends on the alternatives being compared. For example, setting $s_{xy}(r) = 0$ when $|\textbf{Votes}_0(r)| = |\textbf{Votes}_1(r)|$ and $x < y$.
    \end{example}
}

\revision{\noindent In order to define the remaining Unanimity and (Non-)Dictatorship conditions on social welfare functions, we introduce a useful operation for generating pairwise preferences.

    \parsplit

    \begin{definition}
        Let $v \in \{0,e,1\}$ and $N \geq 2$. We define $\Delta v \in \Pair$ as the tuple $(v, v, \dots, v)$, i.e., the tuple of length $N$ with every element equal to $v$.

        \parsplit

        Additionally, for $v \in \{0,1\}$ and $i \in \{1,2,\dots,N\}$, we define $\Delta_i v$ as the tuple $(v,\dots,\neg v,\dots,v)$, i.e., $v$ everywhere except at the $i^{th}$ index, and call the individual $i$ in $\Delta_i$ a \term{deviator}.
    \end{definition}
}

\parsplit

\begin{definition}\label{definition:new-fairness-conditions}
    Let $\A = \AltsSetGeneral$ and $\WelfareSig$ be a social welfare function that satisfies Unrestricted Domain and IIA with pairwise comparison functions $\revision{(s_{xy})_{xy\, \in \Idx}}$. The social welfare function $\Welfare$ satisfies:
    \begin{itemize}
        \item \textbf{Unanimity:} when $\forall\ \revision{xy \in \Idx}$ and $\forall\ \revision{v} \in \2$, we have: $s_\revision{{xy}}(\Delta \revision{v}) = \revision{v}$.
    \end{itemize}

    \begin{itemize}
        \item \textbf{\revision{Individual $i$ is a Dictator:}} when $\forall \revision{xy \in \Idx}$ and $\forall\ (u_1,\dots,u_i,\dots,u_N) \in \Pair$ such that $u_i \in \2$ we have that $s_\revision{xy}(u_1,\dots,u_i,\dots,u_N) = u_i$.
              \noindent If no such individual $i$ exists, we say that $\Welfare$ satisfies \term{Non-Dictatorship}.
    \end{itemize}
    \revision{
        Additionally, for any pair $x, y \in \A$:
        \begin{itemize}
            \item \textbf{Individual $i$ is a Decisive Deviator over $xy$:} when $s_{xy}(\Delta_i 1) = 0$.
        \end{itemize}
    }
\end{definition}

\noindent\revision{Note, a Decisive Deviator over $xy$ is defined for any ordered pair $x, y \in \A$, i.e., even when $xy \notin \Idx$ because an individual may be a Decisive Deviator with respect to the deviator preferencing $x \prec y$ but not when preferencing $y \prec x$, and vice versa.}

\section{Results}\label{section:results}

\subsection{Properties of Social Welfare Functions}\label{subsection:swf-properties}

In this section, we derive a useful lemma about social welfare functions which will be used in our proof of Arrow's Impossibility Theorem. \revision{Namely, that any social welfare function that satisfies Unrestricted Domain, IIA, Unanimity and \revision{Transitivity}, also satisfies each of the following properties defined as follows.}

\revision{\parsplit

    \begin{definition}
        Let $\WelfareSig$ be a social welfare function that satisfies Unrestricted Domain and IIA with pairwise comparison functions $(s_{xy})_{xy\, \in \Idx}$. We say that $\Welfare$ satisfies:

        \begin{itemize}
            \item \textbf{Strictness Preservation:} when $\forall xy \in \Idx$ and $\forall r \in \PairStrict$ we have that $s_{xy}(r) \in \2$.

            \item \textbf{Strict Neutrality:} when $\forall ij, kl \in \Idx: \forall r \in \PairStrict:$
                  \begin{enumerate}
                      \item $s_{ij}(r) = s_{kl}(r)$.

                      \item $s_{ij}(r) = \neg s_{ij}(\neg r)$ \textnormal{(\textit{equivalently,} $s_{ij}(r) = s_{ji}(r)$)}.
                  \end{enumerate}

            \item \textbf{Decisive Deviators are Dictators:} when an individual $i$ being a Decisive Deviator over some $xy \in \Idx$ implies that it is a Dictator.
        \end{itemize}
    \end{definition}
}

\parsplit

\begin{restatable}{lemma}{udproperties}\label{lemma:udproperties}
    \revision{If} $\WelfareSig$ \revision{is} a social welfare function satisfying Unrestricted Domain, IIA, Unanimity \revision{and Transitivity then it satisfies Strictness Preservation, Strict Neutrality and that Decisive Deviators are Dictators}.
\end{restatable}
\begin{proof}
    \normalsize

    \revision{
        \noindent\textbf{Strictness Preservation \& Strict Neutrality:}
        \parsplit

        The fact that social welfare functions satisfying Unrestricted Domain, IIA, Unanimity and Transitivity satisfy Strictness Preservation and Strict Neutrality is well-known (e.g., see~\cite{geanakoplos}). However, for completeness, we prove this solely in terms of the framework of this paper in Appendix~\ref{appendix:proofs}.

        \parsplit

        \noindent\textbf{Decisive Deviators are Dictators:}

        \parsplit

        We assume to the contrary that there is an individual $i \in \{1,2,\dots,N\}$ that is a Decisive Deviator over some $ab$ but that $i$ is not a Dictator, and show that this leads to an aggregate containing a preference cycle (contradicting \revision{Transitivity}).

        \parsplit

        To begin, if $i$ is not a Dictator then $\exists xy \in \Idx$ and some $r = (u_1, \dots, u_i, \dots, u_N) \in \Pair$ with $u_i \in \2$ such that $s_{xy}(r) \neq u_i$. First, consider the case where $xy = A1$.

        \parsplit

        The individual $i$ being a Decisive Deviator over $ab$ implies that for the pairwise comparison $\Delta_i 1 = (1, \dots, 0, \dots 1)$, i.e., $1$ everywhere except at the $i^{\text{th}}$ position, we have that $s_{ab}(\Delta_i 1) = 0$. However, by the strictness of $\Delta_i 1$ and $u_i$, and Strict Neutrality we have that $s_{12}(\Delta_i u_i) = \neg u_i$. Additionally, by Unanimity, $s_{2A}(\Delta \neg u_i) = \neg u_i$.

        \parsplit

        The combination of $s_{12}(\Delta_i u_i) = \neg u_i$ and $s_{2A}(\Delta \neg u_i) = \neg u_i$ yields the following aggregated preferences on $a_1, a_2$ and $a_A$:
        \begin{equation}\label{equation:dd}
            (s_{12}(\Delta_i u_i), s_{2A}(\Delta \neg u_i), s_{A1}(r)) = (\neg u_i, \neg u_i, s_{A1}(r))
        \end{equation}
        However, because $u_i$ is strict and $s_{A1}(r) \neq u_i$, by Lemma~\ref{lemma:condorcet-condition} we have that $(\neg u_i, \neg u_i, s_{A1}(r))$ is a preference cycle on $a_1, a_2$ and $a_A$.

        \parsplit

        Furthermore, by construction, every column in $(\Delta_i u_i, \Delta \neg u_i, r)$ contains each of $u_i$ and $\neg u_i$ so that $(\Delta_i u_i, \Delta \neg u_i, r)$ is a valid profile on $\{a_1, a_2, a_A\}$ (see Note~\ref{note:profile-test}). Thus, by Lemma~\ref{lemma:condorcet-condition}, for any profile $m \in \Prof$ whose pairwise preferences on $12, 2A$ and $A1$ are $(\Delta_i u_i, \Delta \neg u_i, r)$, respectively\footnote{For example, define $m$ by having each individual rank all other alternatives at the bottom of their preference relation and indifferently to one another.}, we have that $\Welfare(m)$ contains a preference cycle on $a_1, a_2$ and $a_A$, contradicting Transitivity.

        \parsplit

        For the remaining cases of $xy \in \Idx$, one just permutes and negates components of the left-hand side of Equation~\ref{equation:dd} in a straightforward manner. For instance, if $1 < x < y < A$, one uses $(s_{1x}(\Delta_i u_i), s_{xy}(r), s_{y1}(\Delta \neg u_i))$, which likewise produces a preference cycle $(\neg u_i, s_{xy}(r), \neg u_i)$ on $1x, xy$ and $y1$ because $s_{y1}(\Delta \neg u_i) = s_{1y}(\Delta \neg u_i) = \neg u_i$ by Strict Neutrality and Unanimity, respectively.\\
    }
\end{proof}

\subsection{Arrow's Impossibility Theorem}\label{subsection:arrow-impossibility}

\revision{In this section we prove Arrow's Impossibility Theorem by generalising D'Antoni's proof in~\cite[Section 4.2]{dantoni} as follows; recalling the equivalence of the below theorem to Arrow's Impossibility Theorem (see Note~\ref{note:equivalence}).}

\parsplit

\begin{theorem}[Arrow's Impossibility Theorem]\label{theorem:arrow-full}
    \revision{If} $\WelfareSig$ is a social welfare function that satisfies \revision{Unrestricted Domain}, IIA, Unanimity and Non-Dictatorship \revision{then it violates Transitivity.}
\end{theorem}
\begin{proof}
    \normalsize

    \revision{We will prove the result by assuming to the contrary that Transitivity holds, and showing that this leads to a contradiction.}

    \parsplit

    Let ${(s_{xy})_{xy\, \in \Idx}}$ be the pairwise comparison functions of $\Welfare$. By Strictness Preservation and Strict Neutrality (Lemma~\ref{lemma:udproperties}), there is some $s: \PairStrict \rightarrow \2$ such that $\forall r \in \PairStrict$ \revision{and $\forall xy \in \Idx: s(r) = s_{xy}(r)$}.

    \parsplit

    \noindent This allows us to form the following two sets.
    \begin{itemize}
        \item $\AggregatesOne{s} \coloneqq \{r \in \PairStrict \mid s(r) = 1\}$, i.e., the set of all pairwise preferences that $s$ aggregates to 1.
    \end{itemize}
    \revision{Then,} $\forall\ r = (u_1,u_2,\dots,u_N) \in \PairStrict$:
    \begin{itemize}
        \item $\VotesOne{r} \coloneqq \{i \in \{1,2,\dots,N\} \mid u_i = 1\}$, i.e., the subset of individuals \textit{voting} 1 in the pairwise preferences of $r$.
    \end{itemize}

    \noindent Moreover, there are two possibilities regarding these sets:
    \begin{enumerate}[label=(\arabic*)]
        \item $\exists\ r \in \AggregatesOne{s}$: $\VotesOne{r} = \{i\}$, i.e., the pairwise preference where the $i^{th}$ individual votes 1 and every other individual votes 0 aggregates to 1.
              \parsplit
        \item $\exists\ r \in \AggregatesOne{s}$: $|\VotesOne{r}| > 1$ minimally, i.e.,  $\nexists\ \revision{q} \in \AggregatesOne{s}$ such that $\VotesOne{\revision{q}} \subset \VotesOne{r}$. In other words, if any of the individuals voting $1$ in $r$ changes their vote, the aggregate becomes $0$.
    \end{enumerate}

    \noindent (1) If $\exists\ r \in \AggregatesOne{s}$ such that $\VotesOne{r} = \{i\}$ then \revision{for $r =$} $(0, \dots, 1, \dots, 0)$ where all arguments of \revision{$r$} are 0 except at the $i^{th}$ position\revision{, we have that $s(r) = 1$.

        \parsplit

        However, $r = \Delta_i 0$, which by Strict Neutrality and Strictness Preservation imply that for any $xy \in \Idx$: $s_{xy}(\Delta_i 1) = s(\Delta_i 1) = s(\neg r) = \neg s(r) = 0$. This implies that $i$ is a Decisive Deviator over $xy$ and hence by Lemma~\ref{lemma:udproperties} is a Dictator, a contradiction.}

    \parsplit

    \noindent (2) Given a minimising $r \in \PairStrict$, we can construct a pair $r', r'' \in \PairStrict$ such that for \revision{$r' = (u'_1, \dots, u'_N) $ and $r'' = (u''_1, \dots, u''_N)$}:
    \begin{itemize}
        \item $\forall\ i \notin \VotesOne{r}$: $\revision{u'_i} = \revision{u''_i} = 1$
        \item $\forall\ j \in \VotesOne{r}$: $\revision{u'_j} = \neg \revision{u''_j}$
    \end{itemize}
    Moreover, $(r, r', r'')$ is profile \revision{on $12, 2A$ and $A1$} (see Table~\ref{table:case-2}). By construction, $\VotesOne{\neg r'} \subset \VotesOne{r}$, which by the minimising condition of $r$ implies that $s(\neg r') = 0$, which by Strict Neutrality implies that $s(r') = 1$. Likewise, $\VotesOne{\neg r''} \subset \VotesOne{r}$ so that $s(r'') = 1$.

    \parsplit

    \revision{Thus, $(s_{12}(r), s_{2A}(r'), s_{A1}(r'')) = (s(r), s(r'), s(r'')) = (1, 1, 1)$, a preference cycle on $a_1, a_2$ and $a_A$.} \revision{Furthermore, by Lemma~\ref{lemma:condorcet-condition}, for any $m \in \Prof$ whose pairwise preferences on $12, 2A$ and $A1$ are $(r, r', r'')$, we have that $\Welfare(m)$ contains a preference cycle on $a_1, a_2$ and $a_A$, contradicting Transitivity.}

    \begin{table}[h]\renewcommand{\arraystretch}{1.1}
        \begin{tabularx}{0.95\textwidth}{|p{2in}|X|X|X|}
            \hline
            \diagbox{Alternatives}{Individual}                          & \textbf{$i$} & $\dots$ & \textbf{$j$} \\\hline
            \textbf{\revision{Row} $r$ \revision{for $a_1$ vs $a_2$}}   & $0$          & $\dots$ & $1$          \\\hline
            \textbf{\revision{Row} $r'$ \revision{for $a_2$ vs $a_A$}}  & $1$          & $\dots$ & $u'_i$       \\\hline
            \textbf{\revision{Row} $r''$ \revision{for $a_A$ vs $a_1$}} & $1$          & $\dots$ & $\neg u'_i$  \\\hline
        \end{tabularx}
        \caption{The matrix $(r, r', r'')$ is a profile because every column has a $0$ and a $1$ \revision{(see Note~\ref{note:profile-test})}.}
        \label{table:case-2}
    \end{table}
\end{proof}

Theorem~\ref{theorem:arrow-full} \revision{is a direct proof of} \revision{Formulation}~\ref{formulation}, demonstrating that Arrow's Impossibility Theorem is equivalent to a statement concerning the necessitation of aggregates \revision{with an intransitive} preference cycle. Specifically, necessitated by the combination of: \revision{Unrestricted Domain}, Unanimity, IIA and Non-Dictatorship. It is specifically a generalisation of Condorcet's Paradox because pairwise majority voting satisfies the same constraints but also aggregates some profiles \revision{with intransitive} preference cycles.

\parsplit

\subsection{Further Insights into Arrow's Impossibility Theorem}\label{section:more}

\revision{In this section, we will provide further insight into the consequences of Arrow's Impossibility Theorem. Specifically, we show that not only is a preference cycle necessitated in some aggregate, but so is a preference cycle spanning all alternatives, which we call a \term{total preference cycle}.

    \parsplit

    Importantly, aggregation to a total preference cycle precludes any global ranking information, i.e., making it impossible to identify a first-ranked candidate, second-ranked candidate, and so on.

    \parsplit

    For example, for alternatives $\A = \{a_1, a_2, a_3, a_4\}$, the following is a total preference cycle on $\A$:
    \begin{equation*}
        a_1 \prec a_2 \prec a_3 \prec a_4 \prec a_1
    \end{equation*}

    \parsplit

    Clearly, no alternative can be considered first, second, third or fourth ranked. However, the same cannot be said for the following complete relation with a non-total preference cycle:
    \begin{equation*}
        a_1 \prec a_2 \prec a_3 \prec a_1 \text{ and } \forall i \neq 4:\ a_i \prec a_4
    \end{equation*}

    \parsplit

    In the second preference cycle, transitivity is violated, but there is still global ranking information, i.e., $a_4$ can be considered first-ranked because it is preferred to all other alternatives.

    \parsplit

    Hence, we proceed to generalise Lemma~\ref{lemma:udproperties} and Theorem~\ref{theorem:arrow-full} by replacing Unrestricted Domain with the condition that a social welfare function has no total preference cycles in its image.

    \parsplit

    Indeed, recall that Lemma~\ref{lemma:udproperties} and Theorem~\ref{theorem:arrow-full} were proven by assuming that Transitivity held, leading to a contradiction. For each result, we constructed a profile with pairwise comparisons $r, r', r'' \in \Pair$ on a triple of alternatives whose aggregate formed a preference cycle.

    \parsplit

    In fact, for each of the above triples $r, r', r''$, the surrounding profile can be constructed out of copies of $r, r', r''$ so that it aggregates to a total preference cycle. This produces the following stronger version of Lemma~\ref{lemma:udproperties} (see Appendix~\ref{appendix:proofs} for the proof).

    \parsplit

    \begin{restatable}{lemma}{grproperties}\label{lemma:grproperties}
        If $\WelfareSig$ is a social welfare function that satisfies Unrestricted Domain, IIA, Unanimity and has no total preference cycles in its image, it satisfies Strictness Preservation, Strict Neutrality and that Decisive Deviators are Dictators.
    \end{restatable}

    Finally, the corresponding strengthening of Theorem~\ref{theorem:arrow-full} constitutes a strengthening of Arrow's Impossibility Theorem.

    \parsplit

    \begin{theorem}\label{theorem:arrow-total-cycle}
        If $\WelfareSig$ is a social welfare function that satisfies \revision{Unrestricted Domain}, IIA, Unanimity and Non-Dictatorship then there exists a profile $m \in \Prof$ such that $\Welfare(m)$ corresponds to a total preference cycle.
    \end{theorem}
    \begin{proof}

        \normalsize
        We will prove the result by assuming to the contrary that $\Welfare$ has no total preference cycles in its image, and showing that this leads to a contradiction.

        \parsplit

        Denoting $\Welfare$'s pairwise comparison functions by $(s_{xy})_{xy\, \in \Idx}$, Strictness Preservation and Strict Neutrality (Lemma~\ref{lemma:grproperties}) imply there exists $s: \PairStrict \rightarrow \2$ such that $\forall r \in \PairStrict$ and $\forall xy \in \Idx: s_{xy}(r) = s(r)$.

        \parsplit

        We then repeat the steps of Theorem~\ref{theorem:arrow-full} to produce pairwise comparisons $(r, r', r'')$ on $a_1, a_2$ and $a_A$ that aggregate to a preference cycle, i.e., where $(s_{12}(r), s_{2A}(r'), s_{A1}(r'')) = (1,1,1)$.  Moreover, because Decisive Deviators are Dictators (Lemma~\ref{lemma:grproperties}), Theorem~\ref{theorem:arrow-full} case (1) is not applicable, which implies we can follow the construction of case (2), and importantly ensure that each of $r, r', r''$ are strict so that $s(r) = s(r') = s(r'') = 1$.

        \parsplit

        Furthermore, we can define a profile $m \in \ProfileStrict$ with row form $(r_{xy})_{xy\, \in \Idx}$, where $r_{1y} \coloneqq r$ for every $1 < y < A$, $r_{A1} \coloneqq r''$ and for every other $xy \in \Idx$, $r_{xy} \coloneqq r'$ (see Lemma~\ref{lemma:extend} for a proof that $m$ is a profile).

        \parsplit

        By IIA, $\Welfare(m) = (s_{xy}(r_{xy}))_{xy\, \in \Idx}$, and $\forall xy \in \Idx$: $s_{xy}(r_{xy}) = s(r_{xy}) = 1$ because $r_{xy} \in \{r, r', r''\}$. Thus, $\Welfare(m) = (1,1,1,\dots,1)$, i.e., corresponding to the total preference cycle $a_1 \succ a_2 \succ a_3 \succ \dots \succ a_A \succ a_1$ (see Proposition~\ref{proposition:spanning-cycles}), contradicting our earlier assumption that $\Welfare$ has no total preference cycles in its image.\\
    \end{proof}
}

\section{Conclusion}\label{section:conclusion}

In this paper, we have \revision{provided a direct proof of a precise sense in which} Arrow's Impossibility Theorem is a generalisation of Condorcet's Paradox on pairwise majority voting. This was achieved by \revision{directly} proving \revision{Formulation}~\ref{formulation}, which states that any social welfare function satisfying all constraints of Arrow’s Impossibility Theorem other than Transitivity necessarily aggregates some profile to a preference cycle. \revision{Specifically, this involved generalising D'Antoni's methodology for proving  \revision{Formulation}~\ref{formulation} for strict preferences~\cite{dantoni}, i.e., allowing indifference between alternatives in individual and aggregate preference relations.}

\parsplit

\revision{Moreover, we leveraged this generalisation to establish that not only are preference cycles guaranteed, but specifically preference cycles that span all available alternatives (Theorem~\ref{theorem:arrow-total-cycle}). This is significant because such preference cycles preclude any global ranking information (i.e., there is no consistent way to determine a first-ranked candidate, second-ranked candidate, and so forth). This result therefore provides further insight into the consequences of Arrow's Impossibility Theorem.}

\parsplit

\revision{In general, the development of methodologies that explore the mechanisms by which} social welfare functions produce preference cycles \revision{may have} broader applicability. \revision{For instance, to variations of Formulation~\ref{formulation} that relax different combinations of Arrow's constraints, including Transitivity~\cite{gibbard-quasi-transitive,blair-acyclic}. These methodologies may also be applicable to situations where the} existence of preference cycles do not \revision{in and of themselves} constitute a contradiction, e.g., \revision{as is the case in} Money Pumps, Dutch Books and Intransitive Games (see~\cite{anand-intransitivity,may-intransitvity,money-pump-gustafsson,dutch-book-hajek} for further examples).

\FloatBarrier
\appendix

\makeatletter

\@removefromreset{theorem}{subsection}
\@addtoreset{theorem}{section}

\makeatother

\renewcommand{\thetheorem}{\thesection.\arabic{theorem}}
\renewcommand{\theproposition}{\thesection.\arabic{theorem}}
\renewcommand{\thedefinition}{\thesection.\arabic{theorem}}
\renewcommand{\thelemma}{\thesection.\arabic{theorem}}
\renewcommand{\thecorollary}{\thesection.\arabic{theorem}}
\renewcommand{\theexample}{\thesection.\arabic{theorem}}
\renewcommand{\thenote}{\thesection.\arabic{theorem}}

\renewcommand{\theHtheorem}{appendix.\thesection.\arabic{theorem}}
\renewcommand{\theHproposition}{appendix.\thesection.\arabic{theorem}}
\renewcommand{\theHdefinition}{appendix.\thesection.\arabic{theorem}}
\renewcommand{\theHlemma}{appendix.\thesection.\arabic{theorem}}
\renewcommand{\theHcorollary}{appendix.\thesection.\arabic{theorem}}
\renewcommand{\theHexample}{appendix.\thesection.\arabic{theorem}}
\renewcommand{\theHnote}{appendix.\thesection.\arabic{theorem}}

\section{Proofs and Supplementary Results}\label{appendix:proofs}

\revision{\begin{proposition}\label{proposition:complete-order-classification}
        A complete binary relation $\preceq$ on a set $\A$ contains a preference cycle if and only if it is intransitive.
    \end{proposition}
    \begin{proof}
        \normalsize

        $(\implies)$ If $\preceq$ has a preference cycle, i.e., $a \preceq b \preceq c \preceq a$ with at least one strict comparison, we can show transitivity is violated by exhaustion. If $a \prec b$ then we have that $b \preceq c$ and $c \preceq a$ but that $b \nprec a$ because we assumed $a \prec b$. Similarly, if $b \prec c$ then $c \preceq a$ and $a \preceq b$ hold but $c \nprec b$. Lastly, if $c \prec a$ then $a \preceq b$ and $b \preceq c$ hold but $a \nprec c$.

        \parsplit

        $(\impliedby)$ If $\preceq$ is not transitive then there exists elements $a,b,c \in \A$ such that $a \preceq b$, $b \preceq c$ but $a \npreceq c$. Thus, if we can prove that $c \prec a$ then we have a preference cycle on $a,b$ and $c$. Indeed, by the completeness of $\preceq$, we must have that one of $a \prec c$, $c \prec a$ or $a \sim c$ holds. However, $a \npreceq c$ implies that neither of $a \prec c$ or $a \sim c$ can hold, thus, the only possibility is that $c \prec a$.\\
    \end{proof}

    \begin{proposition}\label{proposition:spanning-cycles}
        Let $\A = \AltsSetGeneral$ be a set of at least 3 alternatives and $t = (t_{xy})_{xy\, \in \Idx} \in \Pref$. If at least one coordinate of $t$ is $0$ and the rest are in $\{0,\indiff\}$ then $t$ corresponds to the preference cycle with $a_1 \preceq a_2 \preceq a_3 \preceq \dots \preceq a_A \preceq a_1$. Furthermore, if at least one coordinate of $t$ is $1$ and the rest are in $\{1,\indiff\}$ then $t$ corresponds to the preference cycle with $a_1 \succeq a_2 \succeq a_3 \succeq \dots \succeq a_A \succeq a_1$.
    \end{proposition}
    \begin{proof}
        \normalsize

        First consider the case where every $t_{xy} = 0$. This implies that for every $ij \in \Idx$ such that $i < j$: $a_i \prec a_j$ holds, which implies that $a_1 \prec a_2 \prec \dots \prec a_A$. The only other element of $t$ is $t_{A1} = 0$, which implies that $a_A \prec a_1$. Thus, $t$ corresponds to a preference cycle with $a_1 \prec a_2 \prec \dots \prec a_A \prec a_1$. Then allowing some (but not every) $t_{xy} = \indiff$ just transforms certain $\prec$ comparisons to $\sim$, but retaining that $t$ is a preference cycle. Likewise, when each $t_{xy} \in \{e,1\}$, we repeat the above argument, replacing $\prec$ with $\succ$.\\
    \end{proof}}

\test*
\begin{proof}
    \normalsize

    \revision{Let $\preceq$ be the complete relation that $t$ corresponds to. It suffices to show that if conditions 1 and 2 hold for an arbitrary $xy, yz \in \Idx$ then $\preceq$ restricted to $a_x,a_y,a_z$ is transitive, i.e., a weak order. In other words, since transitivity is a ternary condition, a complete relation is transitive if and only if its restriction to every three-element subset is transitive.}

    \parsplit

    \revision{For case 1, we have that $t_{xy} = t_{yz} = t_{zx} = \indiff$} if and only if $t = (\indiff, \indiff, \indiff)$, which corresponds to $a_1 \sim a_2 \sim a_3 \sim a_1$, a valid weak order. \revision{For case 2, assume to the contrary that $\{t_{xy}, t_{yz}, t_{zx}\}$ is one of $\{0\}, \{1\}, \{0,e\}, \{1,e\}$. By Proposition~\ref{proposition:spanning-cycles} this implies that $(t_{xy}, t_{yz}, t_{zx})$ is a preference cycle in $\PrefVar{\{a_x, a_y, a_z\}}$, and hence $t$ has a preference cycle and thus by Proposition~\ref{proposition:complete-order-classification} is intransitive and hence does not correspond to a weak order.}\\
\end{proof}

\noindent\revision{Below is the remainder of the proof of Lemma~\ref{lemma:udproperties}.}

\udproperties*
\begin{proof}
    \normalsize

    \noindent\textbf{Strictness Preservation:}

    \parsplit

    We will prove the result by assuming to the contrary and construct a preference cycle \revision{(contradicting Transitivity).} To begin, assume that $\Welfare$ maps a strict profile to a weak order with indifference, i.e., that $\exists\ r \coloneqq (u_1,u_2\dots,u_N) \in \PairStrict$ and without loss of generality, say $s_{\revision{12}}(r) = \indiff$. Then, $s_{\revision{A1}}(\neg r) = v$ for some $v \in \3$. \revision{This yields the following calculations:}
    \begin{equation*}
        (s_{\revision{12}}(r), s_\revision{2A}(\Delta 0), s_{\revision{A1}}(\neg r)) = (\indiff, 0, v) \quad \text{and} \quad (s_{\revision{12}}(r), s_\revision{2A}(\Delta 1), s_{\revision{A1}}(\neg r)) = (\indiff, 1, v)
    \end{equation*}
    By Lemma~\ref{lemma:condorcet-condition}, for $(\indiff, 0, v)$ to not be a preference cycle \revision{on $a_1, a_2$ and $a_A$}, we require that $v = 1$ but that renders $(\indiff, 1, v)$ a preference cycle \revision{on $a_1, a_2$ and $a_A$}.

    \parsplit

    Finally, because $r$ is strict, $(r, \Delta 0, \neg r)$ and $(r, \Delta 1, \neg r)$ are both profiles \revision{in $\ProfVar{\{a_1,a_2,a_A\}}$ (see Note~\ref{note:profile-test})}. \revision{Hence, for any profile $m \in \Prof$ whose pairwise preferences on $12, 2A$ and $A1$ are $r, r'$ and $r''$, respectively, we have that $\Welfare(m)$ has a preference cycle, contradicting Transitivity.}

    \parsplit

    \noindent\textbf{Strict Neutrality:}

    \parsplit

    \noindent This proof is adapted from~\cite[Theorem 1]{dantoni}. It suffices to prove that \revision{for any $1 < y < A$ that} $\forall\ \revision{r} \in \PairStrict$:
    \begin{enumerate}[label=(\alph*)]
        \item $s_{\revision{1y}}(\revision{r}) = s_{\revision{yA}}(\revision{r}) = \neg s_{\revision{A1}}(\neg \revision{r})$

              \parsplit

        \item $s_{\revision{1y}}(\revision{r}) = s_{\revision{A1}}(\revision{r})$
    \end{enumerate}

    \noindent(a) Firstly, we assume to the contrary that $s_{\revision{yA}}(\revision{r}) \neq \neg s_{\revision{A1}}(\neg \revision{r})$. By Strictness Preservation, $s_{\revision{yA}}(\revision{r})$ and $s_{\revision{A1}}(\neg \revision{r})$ are both in $\2$. Hence, if $s_{\revision{yA}}(\revision{r}) \neq \neg s_{\revision{A1}}(\neg \revision{r})$ then we have that $s_{\revision{yA}}(\revision{r}) = s_{\revision{A1}}(\neg \revision{r})$. Denoting $t \coloneqq s_{\revision{yA}}(\revision{r})$, by the strictness of $\revision{r}$, $(\Delta t, \revision{r}, \neg \revision{r})$ is a profile \revision{on $a_1, a_y, a_A$} (see Note~\ref{note:profile-test}), and so by IIA and Unanimity: $(s_{\revision{1y}}(\Delta t), s_{\revision{yA}}(\revision{r}), s_{\revision{A1}}(\neg \revision{r})) = (t, t, t)$. However, because $t \in \2$: $(t, t, t)$ corresponds to a preference cycle, contradicting \revision{Transitivity}. Thus, we are forced to conclude that $s_{\revision{yA}}(\revision{r}) = \neg s_{\revision{A1}}(\neg \revision{r})$. \revision{Repeating the above argument but switching $1y$ for $yA$, we conclude that $s_{\revision{1y}}(\revision{r}) = \neg s_{\revision{A1}}(\neg \revision{r})$, which in turn equals $s_{\revision{yA}}(\revision{r})$. Hence, $s_{\revision{1y}}(\revision{r}) = s_{\revision{yA}}(\revision{r}) = \neg s_{\revision{A1}}(\neg \revision{r})$, as desired.}

    \parsplit

    \noindent(b) \revision{By the arbitrariness of the initial roles of $1y$, $yA$ and $A1$ in the above proof of (a), we can switch the roles of $yA$ and $A1$ and instead prove that $s_{1y}(r) = s_{A1}(r) = \neg s_{yA}(\neg r)$, i.e., $s_{1y}(r) = s_{A1}(r)$, as desired.}\\
\end{proof}

\revision{\begin{lemma}\label{lemma:extend}
        Let $A = \AltsSetGeneral$ and $N \geq 2$. If for $q, q' \in \PairStrict$ and $q'' \in \Pair$, $(q, q', q'')$ is a profile in $\ProfVar{\{a_1, a_2, a_A\}}$ then the matrix in row form $(q_{xy})_{xy\, \in \Idx}$ defined as follows is a profile in $\Prof$.

        \begin{equation*}
            q_{xy} = \begin{cases*}
                q   & \text{if $x = 1$}   \\
                q'' & \text{if $xy = A1$} \\
                q'  & \text{otherwise.}
            \end{cases*}
        \end{equation*}
    \end{lemma}
    \begin{proof}
        \normalsize
        To verify that $(q_{xy})_{xy\, \in \Idx}$ is a profile, it suffices to check that for all possible $1 \leq i < j < k \leq A$, the following are profiles in $\ProfVar{\{a_i, a_j, a_k\}}$.
        \begin{enumerate}
            \item $(i = 1,\ j = A)$: $(q_{1k}, q_{kA}, q_{A1}) = (q, q', q'')$.
            \item $(1 < i,\ j = A)$: $(q_{ik}, q_{kA}, q_{Ai}) = (q', q', \neg q')$.
            \item $(i = 1,\ j < A)$: $(q_{1k}, q_{kj}, q_{j1}) = (q, q', \neg q)$.
            \item $(1 < i,\ j < A)$: $(q_{ij}, q_{jk}, q_{ki}) = (q', q', \neg q')$.
        \end{enumerate}
        This is because each of the above 3 alternative matrices is a profile only when each of its columns corresponds to a weak order. Then, since the above 4 cases cover every possible triple of alternatives, verifying that each case's columns always correspond to weak orders allows us to conclude the same for $(q_{xy})_{xy\, \in \Idx}$ by Lemma~\ref{lemma:condorcet-condition}.

        \parsplit

        Indeed, we have that case 1 is a profile because $(q, q', q'')$ is a 3 alternative profile by assumption. Then, cases 2, 3 and 4 are profiles because they contain the pair $q$ and $\neg q$, or the pair $q'$ and $\neg q'$, i.e., a strict row and its negation (see Note~\ref{note:profile-test}).\\
    \end{proof}
}

\revision{
    \grproperties*
    \begin{proof}
        \normalsize

        For each property, we assume that $\Welfare$ has no preference cycles in its image, and showing that this leads to a contradiction.

        \parsplit

        Indeed, in our proofs of Strictness Preservation and Strict Neutrality in Lemma~\ref{lemma:udproperties}, every 3 alternative profile $(q, q', q'')$ constructed to aggregate to a preference cycle was strict. This allows us to apply Lemma~\ref{lemma:extend} to produce a profile $m = (q_{xy})_{xy \in \Idx}$, consisting entirely of rows $q, q', q''$. By construction, this yields that each $s_{xy}(q_{xy}) = 1$ for every $xy \in \Idx$. So, by Proposition~\ref{proposition:spanning-cycles}, $\Welfare(m) = (1,\dots,1)$ is a total preference cycle, a contradiction.

        \parsplit

        For the property that Decisive Deviators are Dictators, we assume to the contrary that Non-Dictatorship holds, i.e., some row $r \in \Pair$ violates the dictatorship condition with respect to some pairwise comparison $xy \in \Idx$. While it does not necessarily hold that $xy = A1$, we can prove the result for an equivalent social welfare function where it does hold for $A1$.

        \parsplit

        Specifically, define $\A' = \{a'_1, a'_2, a'_3, \dots, a'_A\}$ by starting with $\A$ and substituting $a_x$ for $a_A$ and $a_y$ for $a_1$, i.e., so that $a'_A \coloneqq a_x$, $a'_x = a_A$, $a'_1 \coloneqq a_y$ and $a'_y = a_1$. Then, we define $\Welfare': \ProfVar{\A'} \rightarrow \PrefVar{\A'}$ to be equivalent to $\Welfare$ i.e., so that it maps the same profiles to the same complete relations, just using different ternary representations. Recall that this can be done by straightforward permutation and negation of the pairwise comparison functions $(s_{xy})_{xy \in \Idx}$ of $\Welfare$.

        \parsplit

        Importantly, if $\Welfare'$ has no total preference cycles in its image, neither does $\Welfare$. However, with $\Welfare'$, we can repeat the construction underlying Equation~\ref{equation:dd} of Lemma~\ref{lemma:udproperties}, producing an $r = (u_1, \dots, u_i, \dots u_N) \in \Pair$ with $u_\in \2$ that violates the dictatorship condition with respect to $A1 \in \A'^{(2)}$, and profile $(\Delta_i u_i, \Delta \neg u_i, r)$ on $a'_1, a'_2$ and $a'_A$ that aggregates to a preference cycle. Finally, because the profile's first two components are strict, we can apply Lemma~\ref{lemma:extend} to produce a profile $m = (q_{xy})_{xy \in \A'^{(2)}}$ such that $q_{A1} = r$ and every other $q_{xy} \in \{\Delta_i u_i, \Delta \neg u_i\}$.  Furthermore, because by construction: $s_{A1}(q_{A1}) \neq u_i$ and $s_{xy}(q_{xy}) = \neg u_i$, otherwise, we have that $\Welfare'(m) = (s_{xy}(q_{xy}))_{xy \in \A'^{(2)}}$ is total preference cycle by Proposition~\ref{proposition:spanning-cycles}.\\
    \end{proof}
}

{\small
    \bibliographystyle{unsrtnat}
    \bibliography{main}
}

\end{document}